\documentclass[12pt,preprint]{aastex}
\newcommand       \be           {\begin{equation}}
\newcommand       \ee           {\end{equation}}
\newcommand       \Angstrom     {\,{\rm \AA}}          
\newcommand       \eV           {\,{\rm eV}\,}
\newcommand       \keV           {\,{\rm keV}\,}
\newcommand       \K            {\,{\rm K}}
\newcommand       \cm           {\,{\rm cm}}
\newcommand       \s            {\,{\rm s}}
\newcommand       \g            {\,{\rm g}}
\newcommand       \erg          {\,{\rm erg}}
\newcommand       \kms          {\,{\rm km \, s}^{-1}}
\newcommand       \pc           {\,{\rm pc}}
\newcommand       \yr           {\,{\rm yr}}
\newcommand       \nH           {n_{\rm H}}

\newcommand       \urad         {u_{\rm rad}}
\newcommand       \uHab         {u_{\rm Hab}^{\rm uv}}
\newcommand       \Jpe          {J_{\rm pe}}
\newcommand       \Jion         {J_{\rm ion}}
\newcommand       \Qabs         {Q_{\rm abs}}

\newcommand       \gtsim        {\gtrsim}
\newcommand       \ltsim        {\lesssim}

\newcommand       \pet          {{\rm pet}}

\shortauthors{Weingartner, Draine, \& Barr}
\shorttitle{X-ray Grain Charging}

\begin{document}

\title{Photoelectric Emission from Dust Grains Exposed to 
Extreme Ultraviolet and X-ray Radiation}

\author{Joseph C. Weingartner\altaffilmark{1}, B. T. 
Draine \altaffilmark{2, 3}, and David K. Barr\altaffilmark{1}}

\altaffiltext{1}{Department of Physics and Astronomy, George Mason University,
MSN 3F3, 4400 University Drive, Fairfax, VA 22030, USA;
joe@physics.gmu.edu, dbarr@physics.gmu.edu}
\altaffiltext{2}{Princeton University Observatory, Peyton Hall,
        Princeton, NJ 08544, USA; draine@astro.princeton.edu}
\altaffiltext{3}{Osservatorio Astrofisico di Arcetri, Largo E Fermi 5, 50125
     Firenze, Italy}

\begin{abstract}

Photoelectric emission from dust plays an important role in grain charging
and gas heating.  To date, detailed models of these processes have focused
primarily on grains exposed to soft radiation fields.  
We provide new estimates of the photoelectric yield for neutral and charged 
carbonaceous and silicate grains, for photon energies exceeding 
$20 \eV$.  We include the ejection of electrons from both the band 
structure of the material and the inner shells of the constituent atoms, 
as well as Auger and secondary electron emission.  We apply the model  
to estimate gas heating rates in planetary nebulae and grain charges
in the outflows of broad absorption line quasars. 
For these applications, secondary emission can be neglected; the combined 
effect of inner shell and Auger emission is small, though not always 
negligible.  
Finally, we investigate
the survivability of dust entrained in quasar outflows.  The lack of 
nuclear reddening in broad absorption line quasars may be explained by 
sputtering of grains in the outflows.

\end{abstract}

\keywords{ISM: dust}

\section{\label{sec:intro} Introduction}

Cosmic grains are subjected to electrical charging processes and the 
resulting non-zero charge can have important astrophysical consequences.
In Galactic \ion{H}{1} regions, the grain charge is primarily determined
by a balance between starlight-induced photoelectric emission and the 
collisional capture of electrons from the gas.  Furthermore, in these 
regions the gas heating is dominated by photoelectric emission from dust.  
Thus, 
there have been numerous detailed studies of UV-induced photoelectric emission 
(Spitzer 1948; Watson 1972; de Jong 1977; Draine 1978; Tielens \& 
Hollenbach 1985; Bakes \& Tielens 1994; Weingartner \& Draine 2001b).

Photoelectric 
emission from grains exposed to extreme ultraviolet (EUV) and X-ray 
radiation has received much less attention.  Dwek \& Smith (1996)
modelled, in detail, the dust and gas heating for neutral grains
exposed to high-energy radiation, but did not address the grain charging.

Charging by high-energy photons can be very important for grains immersed
in hot plasma and exposed to a hard radiation field.  The grain potential
is limited by the highest photon energy present in the incident radiation;
as this increases, the grain can reach higher potentials.  The efficiency of
the gas heating decreases as the grain charge increases, since the 
photoelectrons have to climb out of the potential well.  

Hard radiation fields can be found, e.g., in planetary nebulae and 
active galactic nuclei (AGNs).  If the 
gas contains dust, then a fraction of the incident radiation will be 
processed into infrared (IR) radiation, i.e., thermal dust emission.  Dwek
\& Smith (1996) were motivated by the prospect of learning about these 
environments through the analysis of the IR emission.  Ferland et 
al.~(2002) stressed 
that the X-ray and IR spectra in AGNs may be correlated as a result of the 
interaction between hot grains and gas.  Detailed modelling of these 
regions will only be possible once the grain charging is properly modelled,
since the grain and gas heating rates depend on the charging.  

In a previous paper (Weingartner \& Draine 2001b, hereafter WD01), we 
modelled in detail the photoelectric emission from grains exposed to UV
radiation; this is briefly summarized in \S \ref{sec:summ}.
In \S \S \ref{sec:pe} through \ref{sec:htg}, 
we extend the WD01 model to include EUV and X-ray radiation.  
Applications to planetary nebulae and quasar outflows are described in 
\S \ref{sec:PN} and \S \ref{sec:quasar}, respectively.

\section{\label{sec:summ} Photoelectric Emission Induced by Low-Energy Photons}

As a starting point for this study, we will adopt the WD01 photoelectric
emission
model.  Here, we briefly discuss some of the important features of this
model and mention two minor modifications.  

The photolectric emission rate depends on the absorption cross section and the 
photoelectric yield $Y$ (i.e., the probability that an electron will be 
emitted following the absorption of a photon), both of which depend on 
the photon energy $h\nu$.  

WD01 assumed spherical grains, for which the absorption cross section is 
$\Qabs \pi a^2$, where $a$ is the grain radius and $\Qabs$ is the 
absorption efficiency factor.  They used Mie theory to compute $\Qabs$
and adopted dielectric functions from Li \& Draine (2001) and Weingartner 
\& Draine (2001).  Here, we will use Mie theory to compute $\Qabs$ 
when $x \equiv 2\pi a/ \lambda < 2 \times 10^4$ ($\lambda$ is the 
wavelength of the radiation) and anomalous diffraction 
theory when $x > 2 \times 10^4$ (Draine 2003).  We also will use 
somewhat newer dielectric functions (Draine 2003).

The WD01 prescription for determining the threshold photon energy for 
photoelectric 
emission, $h\nu_{\rm pet}$, is discussed in their \S \S 2.2 and 2.3.1
and equations 2 through 7.  Here, we will adopt a different expression for
the minimum energy that an electron must have to escape a 
negatively-charged grain, $E_{\rm min}$.  Instead of equation 7 from 
WD01, we adopt equation 1 from van Hoof et al.~(2004).\footnote{Since
van Hoof et al.~(2004) define $E_{\rm min}$ as the negative of our  
$E_{\rm min}$, the minus sign in their equation 1 is absent here.}
Thus, $h\nu_{\rm pet}$ for photoelectric emission from the 
band structure is taken to be
\be
h\nu_{\rm pet}(Z, a) = \cases{
{\rm IP}_V(Z, a) &, $Z \ge -1$\cr
{\rm IP}_V(Z, a) + E_{\rm min}(Z, a) &, $Z < -1$\cr
}
\ee
with the ``valence band ionization potential''
\be
\label{eq:IP_V}
{\rm IP}_V(Z, a) = W + \left( Z + \frac{1}{2} \right) \frac{e^2}{a} + 
\left( Z + 2 \right) \frac{e^2}{a} \frac{0.3 \Angstrom}{a}
\ee
and 
\be
\label{eq:E_min}
E_{\rm min} = \cases{
0 &, $Z \ge -1$\cr
\theta_{\nu}(\nu = |Z+1|) \left[ 1 - 0.3(a/10 \Angstrom)^{-0.45}
|Z+1|^{-0.26} \right] &, $Z < -1$\cr
}~~~;
\ee
$W$ is the work function and $\theta_{\nu}$ is defined in equation 2.4
of Draine \& Sutin (1987).  We retain the WD01 estimates of $W = 4.4 \eV$
($8 \eV$) for carbonaceous (silicate) grains.

WD01 adopted the following expression for the photoelectric yield of a grain
with charge $Ze$ (their equation 12; $e$ is the proton charge):
\be
Y(h\nu, Z, a) = y_2(h\nu, Z, a) \times
\min \left[ y_0(\Theta) y_1(a, h\nu), 1 \right]~~~, 
\label{eq:yeqn}
\ee
where
\be
\label{eq:Theta}
\Theta = \cases{
h\nu - h\nu_\pet + (Z+1)e^2/a &, $Z \ge 0$\cr
h\nu - h\nu_\pet &, $Z < 0$\cr
}
\ee
(WD01 equation 9).  For a bulk solid, $Y(h\nu) = y_0(\Theta = h\nu -W)$.
WD01 present approximations, derived
from laboratory measurements, for $y_0(\Theta)$ for carbonaceous and silicate 
grains (their equations 16 and 17, respectively).  

We adopt the size-dependent yield enhancement factor $y_1$ used by WD01:
\be
\label{eq:y1}
y_1(a,h\nu) = \left(\frac{\beta}{\alpha}\right)^2
                   \frac{\alpha^2-2\alpha+2-2\exp(-\alpha)}
                        {\beta^2-2\beta+2-2\exp(-\beta)}
\ee
where $\beta\equiv a/l_a$ and $\alpha\equiv a/l_a+a/l_e$.
It depends on the 
photon attenuation length, $l_a$ (the $e$-folding length for the decrease
of radiation intensity as it propagates into the material) and the 
electron escape length, $l_e$ (roughly the distance that the electron travels
in the material before losing its energy).  The electron escape length 
depends on the energy of the excited photoelectron; WD01 adopted 
$l_e = 10 \Angstrom$ in all cases, which is a reasonable approximation for
the low-energy electrons excited by visible and UV radiation.  

Finally, the factor $y_2(h\nu, Z, a)$ accounts 
for the fact that not all of the electrons that breach the surface barrier 
have sufficient energy to escape to infinity. WD01 assumed a parabolic 
electron energy distribution:
\be
\label{eq:parab}
f_E^0(E) = \frac{6 (E - E_{\rm low}) (E_{\rm high} -E)}
{(E_{\rm high} - E_{\rm low})^3}~~~~~~~~~~~,~~~~~E_{\rm low}
\leq E \leq E_{\rm high}~~~,
\ee
where $f_E^0(E) dE$ gives the fraction of attempting electrons with
energy (with respect to infinity) between $E$ and $E + dE$.  When $Z<0$,
$E_{\rm low} = E_{\rm min}$ and $E_{\rm high} =
E_{\rm min} + h\nu - h\nu_\pet$; when $Z \ge 0$, $E_{\rm low} = -(Z+1)
e^2/a$ and $E_{\rm high} = h\nu - h\nu_\pet$.  
The fraction of attempting electrons that escape to infinity is given by
\be
\label{eq:y2}
y_2(h\nu, Z, a) =
\cases{
\int_0^{E_{\rm high}} dE f_E^0(E) =
E_{\rm high}^2 (E_{\rm high} - 3 E_{\rm low})/(E_{\rm high} -
E_{\rm low})^3~~~&, $Z\geq 0$\cr
1               &, $Z < 0$\cr
}~~~.
\ee

\section{\label{sec:pe} Photoelectric Emission Induced by High-Energy Photons}

Photoelectric 
emission induced by high-energy photons differs from that induced 
by low-energy photons in the following ways:

\noindent 
1.  Whereas low-energy photons can only excite electrons from the band 
structure of the solid, high-energy photons can also excite inner shell
electrons.  Except very near the absorption edge (see, e.g., Draine
2003) the absorption cross section from inner shell electrons will
be essentially identical to that for isolated atoms.  Dwek \& Smith (1996) 
find that the transition between band-like and atomic-like absorption 
typically occurs for photon energies around $50 \eV$.  However, 
laboratory-measured photoelectron yields are only available for 
$h\nu$ up to $\approx 20 \eV$.  

\noindent
2.  A sufficiently energetic photon can excite a
photoelectron from atomic shells other than the highest occupied shell
(e.g., from the 1s shell in C if $h\nu > 291 \eV$).  The hole left by the 
photoelectron can then be filled in a radiationless transition in which 
a second electron fills the vacancy produced by the photoionization, with 
the excess binding energy going into kinetic energy of a third electron, 
which then leaves the atom (the Auger effect).
The Auger electron can then possibly leave the grain.  
This process can give rise to a cascade of secondary Auger transitions 
that fill holes produced in previous Auger transitions.  
Thus, more than one electron can be 
emitted from a grain following the absorption of a high-energy photon.

\noindent
3.  Following the absorption of a high-energy photon, the photolectrons and 
Auger electrons might have enough energy to excite secondary electrons from 
the grain.  

In the following sections, we will discuss the emission of primary 
photoelectrons, Auger electrons, and secondary electrons.

\section{Emission of Primary Photoelectrons}

\subsection{Bulk Yields}

\subsubsection{Physical Model \label{sec:emission_model}}

In the case of high-energy radiation, we assume atomic-like absorption 
and estimate the bulk yield of photoelectrons ejected from each 
shell of each atomic constituent of the grain:  
$y_0(i, s; \Theta_{i,s})$ is the probability that a primary photoelectron 
excited from shell $s$ of element $i$ escapes the bulk solid following the 
absorption of a photon of energy $h\nu = \Theta_{i,s} + I_{i,s}$, where
$I_{i,s}$ is the ionization energy of shell $s$ of element $i$.  

To estimate the bulk photoelectron 
yields, we adopt a semi-infinite slab geometry and assume that a 
photoelectron escapes the solid 
with probability $0.5 \, \exp(-z/l_e)$, where $z$ is the perpendicular distance
between the surface and the point of excitation.  The factor 0.5 accounts
for the fact that only half of the photoelectrons will travel towards the
surface of the solid, if they are emitted isotropically.  
For simplicity, we ignore refraction and the fact that the reflectivity
varies with the angle of incidence, $\theta$.  For an isotropic incident
radiation field, 
\be
\label{eq:y0}
y_0(i,s; \Theta_{i,s}) 
= n_i \sigma_{i,s} l_a \int_0^{\pi/2} d\theta \sin \theta
\cos \theta \int_0^{\infty} \frac{dx}{l_a} \exp \left[ -x \left( 
\frac{1}{l_a} + \frac{\cos \theta}{l_e} \right) \right]
= n_i \sigma_{i,s} l_e \left[ 1 - \frac{l_e}{l_a} \ln \left( 1 + 
\frac{l_a}{l_e} \right) \right]~~~;
\ee
$x$ is the distance along an incident ray from the surface and 
$n_i \sigma_{i,s} l_a$ is the probability that absorption is by shell $s$
of element $i$.  In evaluating $\sigma_{i,s} l_a$, the photon energy 
$h\nu = \Theta_{i,s} + I_{i, s}$; 
in evaluating $l_e$, the initial energy of the excited 
photoelectron $E_e = h\nu - I_{i, s} = \Theta_{i,s}$.  
To find the bulk yield of photoelectrons ejected from the band structure,
$y_0({\rm band}, \Theta_{\rm band})$, we sum $y_0(i,s)$ over all of 
the shells that comprise the band, except that we take 
$E_e = \Theta_{\rm band}$ and $h\nu = \Theta_{\rm band} + W$.  
We take photoionization cross sections from Verner \& Yakovlev (1995) and
Verner et al.~(1996), making use of the FORTRAN routine
{\it phfit2}.\footnote{The subroutine {\it phfit2} was written by
D.~A.~Verner and is available at
http://www.pa.uky.edu/$\sim$verner/fortran.html.}

The photon absorption length $l_a=\lambda/(4\pi \, {\rm Im} \, m)$,
where $\lambda$ is the wavelength {\it in vacuo} and
$m(\lambda)$ is the complex refractive index, can be approximated by
\be
\label{eq:l_a1}
l_a^{-1}(h\nu) = \sum_{i,s}  n_i \sigma_{i,s}(h\nu)~~~.
\ee
For consistency with our atomistic approach to photoelectric emission, 
we will use the approximation when $h\nu > 20 \eV$.

\subsubsection{Electron Escape Length}

WD01 took $l_e = 10 \Angstrom$ for $h\nu \ltsim 20 \eV$, in approximate
agreement with experiments on C (Martin et al.~1987) and SiO$_2$ 
(McFeely et al.~1990) films.  For high initial energies $E_e$, the electron 
escape length is roughly given by (Draine \& Salpeter 1979)
\be
\label{eq:l_e_high_energy}
l_e(E_e) \approx 300 \Angstrom \left( \frac{\rho}{\g \cm^{-3}}
\right)^{-0.85} \left( \frac{E_e}{\keV} \right)^{1.5}~~~,~~~300 \eV < E_e
< 1 \, {\rm MeV}~~~,
\ee
where $\rho$ is the density of the material.  Extending equation
(\ref{eq:l_e_high_energy}) below $300 \eV$, we would find $l_e = 10 
\Angstrom$ at $E_e = 164 \eV$ for carbonaceous grains (assuming the ideal
graphite density of $\rho = 2.24 \g \cm^{-3}$) and at 
$E_e = 211 \eV$ for silicates (assuming $\rho = 3.5 \g \cm^{-3}$).  
The upturn in $l_e$ at $\sim 200 \eV$ for C is in rough agreement with 
the results of Martin et al.~(1987), who considered $E_e$ up to $1 \keV$.
Thus, we adopt
\be
\label{eq:l_e_carb}
l_e(E_e; \, {\rm carbonaceous}) = \cases{10 \Angstrom &, $E_e \le 164 \eV$\cr
4.78 \times 10^{-3} \Angstrom (E_e/\eV)^{1.5} &, $E_e > 164 \eV$\cr}
\ee
and 
\be
\label{eq:l_e_sil}
l_e(E_e; \, {\rm silicate}) = \cases{10 \Angstrom &, $E_e \le 211 \eV$\cr
3.27 \times 10^{-3} \Angstrom (E_e/\eV)^{1.5} &, $E_e > 211 \eV$\cr}~~~.
\ee

\subsubsection{Carbonaceous Grains \label{sec:bulk_yield_gra}}

\begin{figure}
\epsscale{1.00}
\plotone{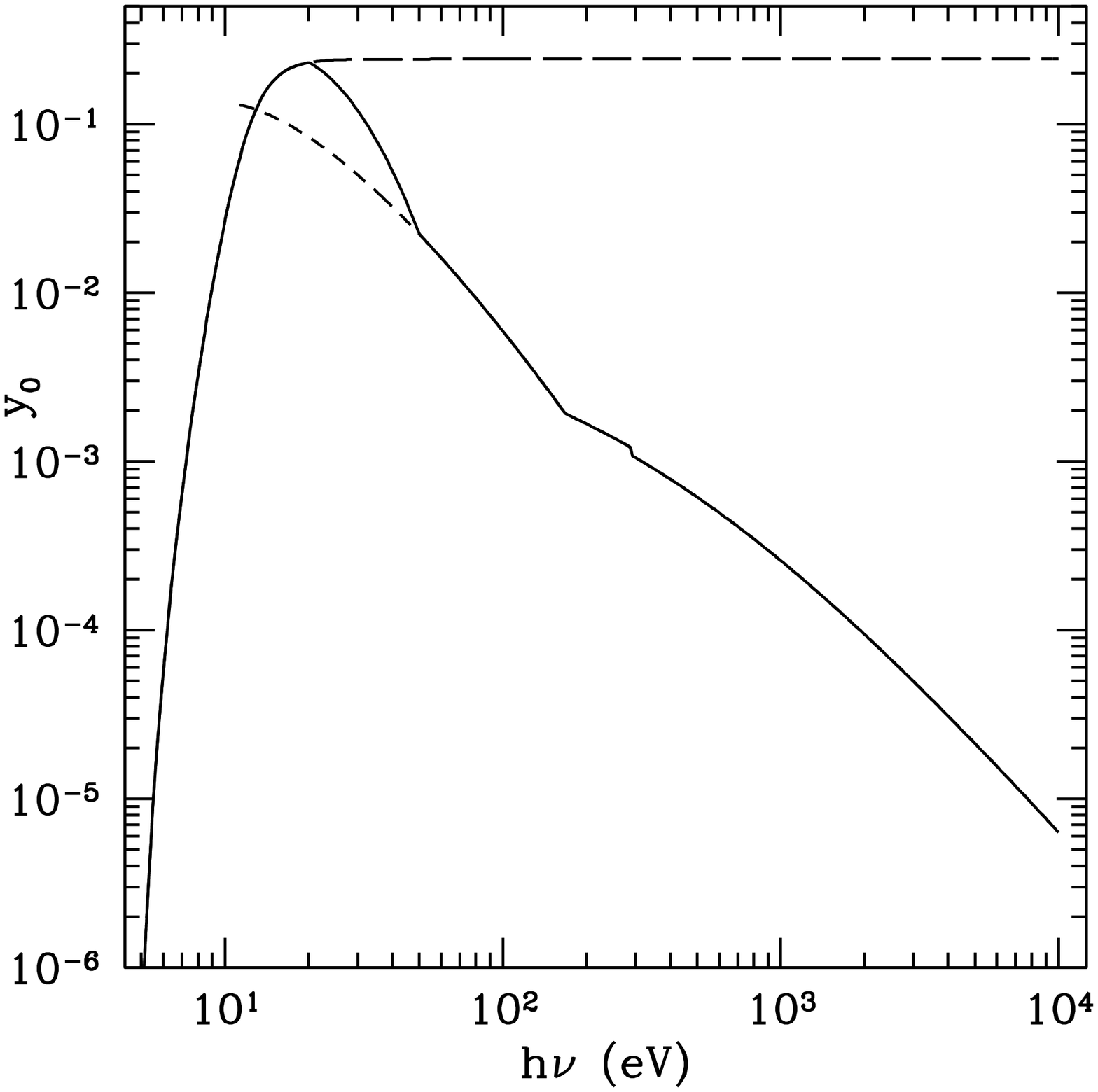}
\caption{
\label{fig:y0_gra_band}
Bulk yield for carbonaceous dust.  Long-dashed curve:  WD01 yield; 
short-dashed: yield computed with eq.~(\ref{eq:y0}); solid:  our adopted yield 
        }
\end{figure}

We adopt the
ideal graphite density of $2.24 \g \cm^{-3}$; thus, the C atom
number density $n_{\rm C} = 1.12 \times 10^{23} \cm^{-3}$.  There are 
strong correlations between the 2s and 2p shells, with no jump in the 
photoionization cross section at the 2s threshold (Verner et al.~1996).
Thus, we treat (2s + 2p) as a single shell, with ionization energy 
equal to that of the 2p shell $= 11.26 \eV$.  (See Table \ref{tab:IE} for
the ionization energies of relevant elements.)  The bulk yield computed
using equation (\ref{eq:y0}) is displayed as the short-dashed curve in 
Figure \ref{fig:y0_gra_band}.  Since the (2s + 2p) shell produces the 
band structure, this high-energy yield curve should connect continuously
with the low-energy curve from WD01 (their equation 16), which is 
displayed as the long-dashed curve in Figure \ref{fig:y0_gra_band}.
To enforce this continuity, we adopt the yield computed here when 
$h\nu > 50 \eV$, the WD01 yield when $h\nu < 20 \eV$, and interpolate
between these when $20 \eV < h\nu < 50 \eV$.  The result is displayed as
the solid curve in Figure \ref{fig:y0_gra_band}.

\subsubsection{Silicate Grains \label{sec:bulk_yield_sil}}

For silicate grains, we adopt a stoichiometry approximating MgFeSiO$_4$
and a density of $3.5 \g \cm^{-3}$, intermediate between the values for 
crystalline forsterite (Mg$_2$SiO$_4$, $3.21 \g \cm^{-3}$) and fayalite
(Fe$_2$SiO$_4$, $4.39 \g \cm^{-3}$).  The atomic number densities are thus
$n_{\rm Mg} = n_{\rm Fe} = n_{\rm Si} = 1.22 \times 10^{22} \cm^{-3}$ and
$n_{\rm O} = 4.88 \times 10^{22} \cm^{-3}$.

\begin{figure}
\epsscale{1.00}
\plotone{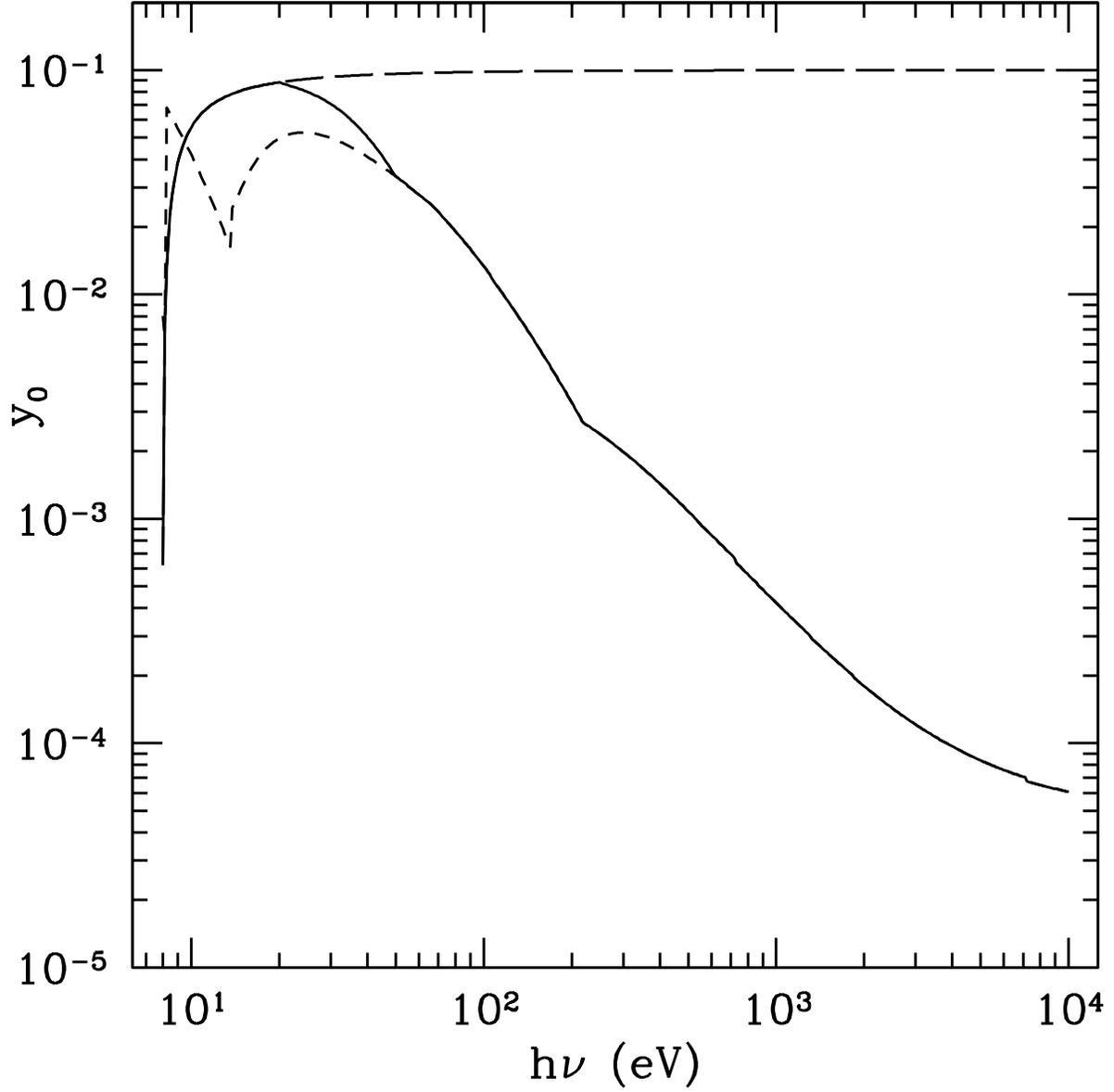}
\caption{
\label{fig:y0_sil_band}
Bulk yield for silicate dust.  Long-dashed curve:  WD01 yield; short-dashed:
yield computed with eq.~(\ref{eq:y0}); solid:  our adopted yield
        }
\end{figure}

As with C (2s + 2p), O (2s + 2p), Si (3s + 3p), and Fe (3d + 4s) do not
display jumps in the photoionization cross section at the threshold 
energy of the deeper shell.  Thus, we treat each of these pairs as 
a single shell, with ionization energy equal to that of the lower-energy 
shell.  The O 2s and 2p, Si 3s and 3p, Mg 3s, and Fe 3d and 4s
shells are considered to comprise the silicate band structure.  As with
graphite, we adopt the WD01 yield when $h\nu < 20 \eV$, the yield computed
here when $h\nu > 50 \eV$, and interpolate between these when 
$20 \eV < h\nu < 50 \eV$.  The resulting yield is displayed in Figure
\ref{fig:y0_sil_band}.

\subsection{Size-Dependent Yield \label{sec:size_dep}}

For photoelectric 
emission from the band structure of carbonaceous and silicate grains,
we compute $y_1(h\nu, a)$ and $y_2(h\nu, Z, a)$ using the WD01 prescription,
as modified in \S \ref{sec:summ} above (eqs.~\ref{eq:y1} and \ref{eq:y2}).
We estimate $y_0(\Theta_{\rm band})$ as discussed in \S \S
\ref{sec:bulk_yield_gra} and \ref{sec:bulk_yield_sil}; $\Theta_{\rm band}$
is computed using equation (\ref{eq:Theta}).  Equation (\ref{eq:yeqn}) is
employed to compute the yield, except that the 1 in the min function is
replaced by the probability $P_{\rm band}$ that the photon absorption occurs 
in the band structure, rather than in an inner shell.  We estimate 
\be
P_{\rm band}(h\nu) \approx l_a(h\nu) \sum_{i,s}^{\rm band} 
n_i \sigma_{i,s}(h\nu)~~~,
\ee
where the sum is over all shells that are taken to comprise the band 
structure (see \S \S  \ref{sec:bulk_yield_gra} and 
\ref{sec:bulk_yield_sil}).  

The yield for photoelectric 
emission from the inner shells is computed in exactly 
the same way, except that the work function is replaced by the appropriate
ionization energy when computing the photoelectric threshold energy, 
$h\nu_{\rm pet}$ (which is needed to evaluate $\Theta_{i,s}$; see 
eq.~\ref{eq:Theta}).  Also, $P_{\rm band}$ is replaced with 
$P_{i,s} = l_a n_i \sigma_{i,s}$.  

Very small carbonaceous grains are taken to be polycyclic aromatic 
hydrocarbons (PAHs).  For these, we expect $Y \rightarrow 1$ as 
$h\nu \rightarrow \infty$ for the electrons associated with the ``band
structure''.  This applies by construction for the WD01 yields, but does
not apply to the yields derived here.  Thus, when $a \le 6 \Angstrom$
(corresponding to $\approx 100$ C atoms; see eq.~1 in WD01), we employ 
$y_0$ from WD01.  When $6 \Angstrom < a \le 13 \Angstrom$, we take
\be
\label{eq:y_band_PAH}
y_{0, {\rm band}} = 
y_{0, {\rm band}}({\rm WD01}) \frac{13 \Angstrom - a}{7 \Angstrom} + 
y_{0, {\rm band}}({\rm this \ work}) \frac{a - 6 \Angstrom}{7 \Angstrom}~~~.
\ee

\section{Emission of Auger Electrons \label{sec:Auger}}

Suppose a primary photoelectron is ejected from shell $s$
of element $i$.  There are a number of Auger transitions that can then 
occur; we will denote them with index $j$.  Adopting the model for electron
escape from \S \ref{sec:emission_model}, the bulk yield for the $j$-th
Auger transition is
\be
\label{eq:y0_A}
y_{0, A}(i,s,j; \Theta_{A;i,s}) = p_{A,i,s,j} n_i \sigma_{i,s} l_e \left[ 1 - 
\frac{l_e}{l_a} \ln \left( 1 + \frac{l_a}{l_e} \right) \right]~~~,
\ee
where $p_{A,i,s,j}$ is the average number of electrons ejected from the
atom via Auger transition $j$, photon energy $h\nu = \Theta_{A; i,s} + 
I_{i, s}$ is used in evaluating $\sigma_{i,s}$ and $l_a$, and the energy
$E_{A, i,s,j}$ of the Auger electron is used when evaluating $l_e$.
We adopt the values of $p_{A,i,s,j}$ and $E_{A, i,s,j}$ given in 
Tables 4.1 and 4.2 of Dwek \& Smith (1996).  Note that the Auger yield is
the average number of electrons emitted by the grain via the given
Auger transition, rather than the probability of electron emission, 
and may exceed unity.  

For all grain sizes and charges, we take the threshold photon energy equal to
$I_{i,s}$, even though it should really be somewhat less than
this, due to the presence of the band structure.  
We use equation (\ref{eq:yeqn}) to evaluate the
Auger electron yield $Y_{A, j}$, except that the 1 in the min function 
is replaced with $Y_{A,i,s,j;{\rm max}} = P_{i,s} p_{A,i,s,j}$.
We also adopt
somewhat different values of the input energies than for the
case of primary photoelectrons.  In evaluating $y_{0, A}$, we take
$\Theta_{A;i,s} = h\nu - I_{i,s}$, regardless of grain size and charge.
In computing $y_1(h\nu, a)$, we take
$E_e = E_{A,i,s,j}$ (when evaluating $l_e$).  We use equation (\ref{eq:y2}) to
compute $y_2(h\nu, Z, a)$ with $E_{\rm high} = E_{A, i,s,j} - (Z+1) e^2/a$;
$E_{\rm low} = E_{\rm min}$ when $Z < 0$ and $E_{\rm low} = -(Z+1) e^2/a$
when $Z \ge 0$.  For Auger electrons, it can occur that
$E_{\rm high} < 0$; in this case, we set $y_2 = 0$.

\section{Emission of Secondary Electrons}

The secondary electron yield is defined as the average number of secondary 
electrons emitted per absorbed photon; it may exceed unity.  
This yield can be expressed as a sum of partial yields (denoted with index
$k$), with a term for each type of process that can excite a secondary 
electron (namely, primary emission from the band structure and inner
shells and Auger emission):
\be
Y_{\rm sec}(h\nu, Z, a) = \sum_k Y_{{\rm sec} \, , k}(h\nu, Z, a)~~~.
\ee
Since secondary electron energies are generally 
low, only those excited close to the surface have significant
probability of escape.  Thus, we take $Y_{{\rm sec} \, , k} = 0$ if 
$Y_k = 0$; otherwise, we approximate
\be
Y_{{\rm sec} \, , k}(h\nu, Z, a) = Y_k(h\nu, Z, a)
\frac{y_{2,k}^{\rm sec}(h\nu, Z, a)}{y_{2,k}(h\nu, Z, a)}
\frac{E_{e,_k}}{\epsilon}
\frac{l_e^{-1}(E_{e,k})}{a^{-1} + (10 \Angstrom)^{-1}}
\ee
where $E_{e,k} = h\nu -W$ for primary electrons excited from the band
structure, $h\nu - I_{i,s}$ for primary electrons excited from inner 
shell ($i$, $s$), and $E_{A,i,s,j}$ for an Auger electron; 
$\epsilon$ is the average energy loss by the exciting electron per secondary 
electron created in the solid.  Draine \& Salpeter (1979) estimate
that $\epsilon = 117 \eV$ for graphite and $155 \eV$ for lunar dust; we
will adopt these values for carbonaceous and silicate grains.  The grain size 
$a$ is included in the last factor to account for the fact that the exciting
electron may escape the grain before producing secondary electrons.

For the secondary electron energy distribution, we adopt
\be
\label{eq:f_sec}
f_E^0(E) = \frac{\alpha \tilde{E}^{-2} (E - E_{\rm low})}
{\left[ 1 + \tilde{E}^{-2} (E - E_{\rm low})^2 \right]^{3/2}}
\ee
(Draine \& Salpeter 1979) with $\tilde{E}^2 = 8 \eV^2$ and normalization 
factor 
\be
\alpha = \left\{ 1 - \left[ 1 + \tilde{E}^{-2} \left( E_{\rm high} - 
E_{\rm low} \right)^2 \right]^{-1/2} \right\}^{-1}~~~.
\ee
Thus, 
\be
y_{2, k}^{\rm sec}(h\nu, Z, a) =
\cases{
\alpha \left\{ \left[ 1 + \tilde{E}^{-2} E_{\rm low}^2 \right]^{-1/2} - 
\left[ 1 + \tilde{E}^{-2} \left( E_{\rm high} - E_{\rm low} \right)^2 
\right]^{-1/2} \right\}~~~&, $Z\geq 0$\cr
1               &, $Z < 0$\cr
}
\ee
when $E_{\rm high} > 0$; otherwise, $y_{2, k}^{\rm sec} = 0$.  
For secondary electrons 
excited by a primary photoelectron, we take $E_{\rm low}$
and $E_{\rm high}$ to be the same as for the primary; likewise for 
those excited by Auger electrons.
Consequently, the threshold photon energies for the emission of secondary
electrons are equal to those for the electrons that excite them.  

\begin{figure}
\epsscale{1.00}
\plotone{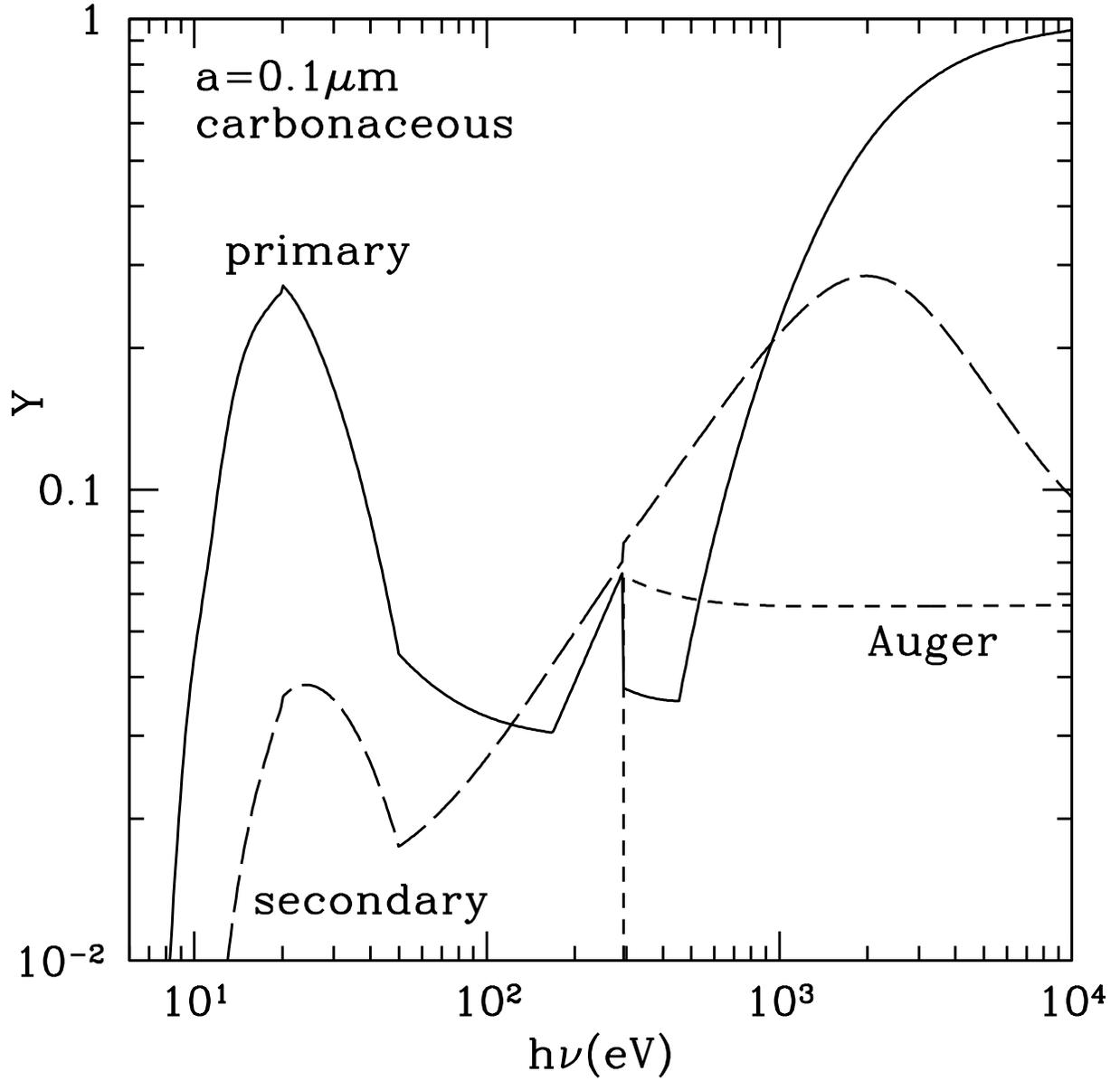}
\caption{
\label{fig:y_gra_0.1}
The yields of primary, Auger, and secondary electrons for uncharged
carbonaceous grains with $a=0.1 \micron$.  
        }
\end{figure}

\begin{figure}
\epsscale{1.00}
\plotone{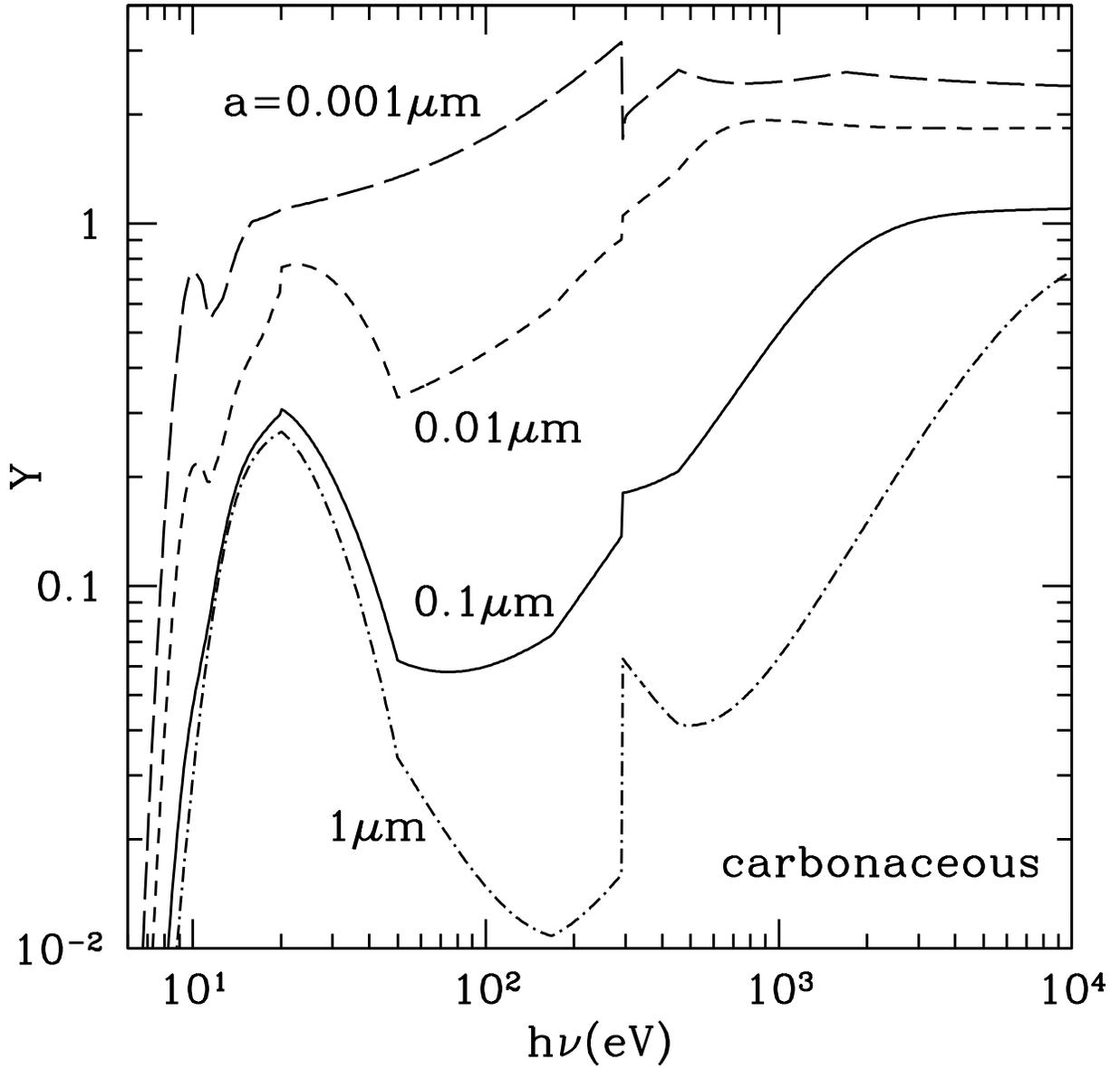}
\caption{
\label{fig:y_tot_gra}
The total yield (primary plus Auger plus secondary) for uncharged
carbonaceous grains of four different sizes, as labelled.  
        }
\end{figure}

\begin{figure}
\epsscale{1.00}
\plotone{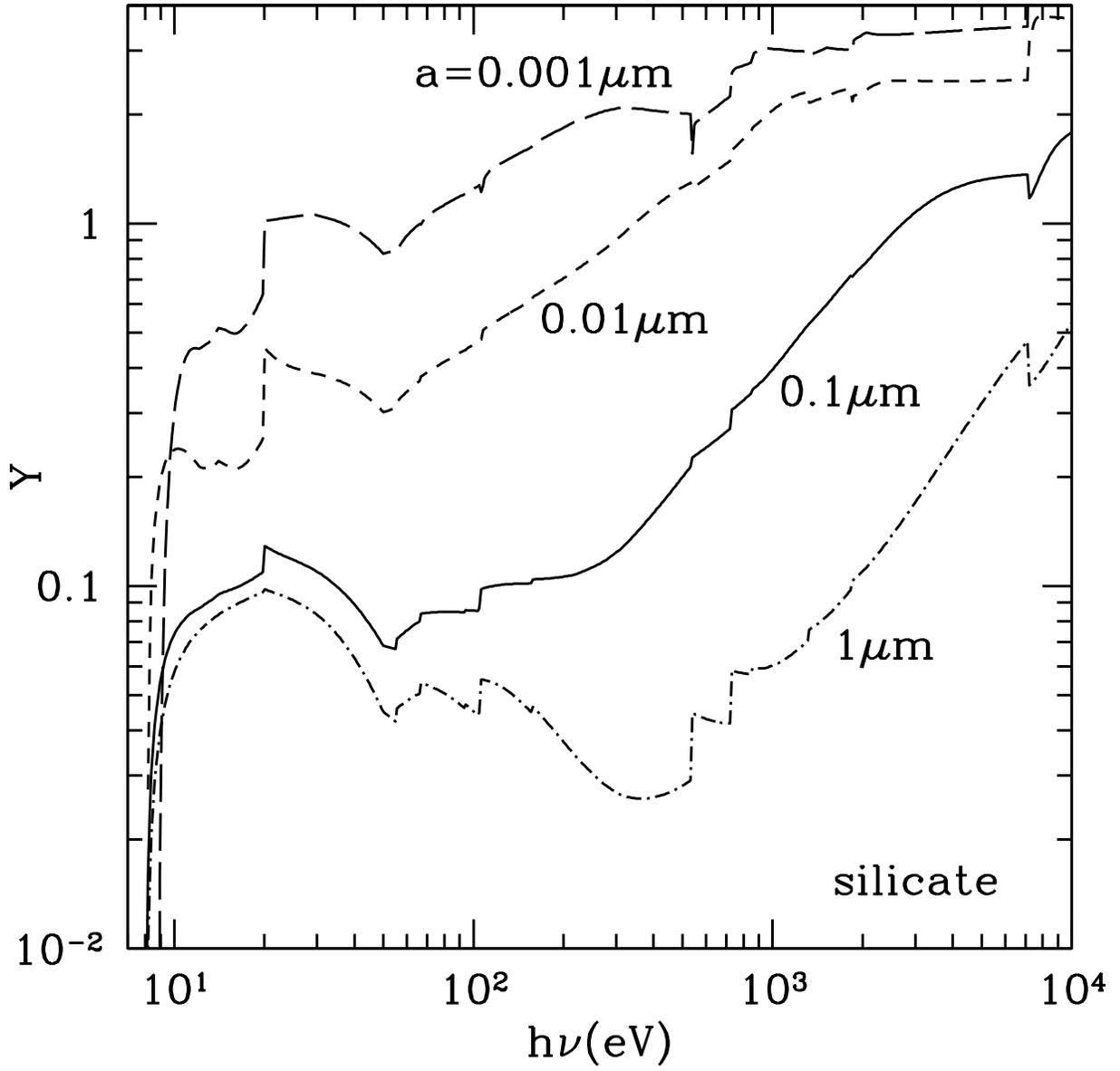}
\caption{
\label{fig:y_tot_sil}
Same as Fig.~\ref{fig:y_tot_gra}, except for uncharged silicate grains.
        }
\end{figure}

Figure \ref{fig:y_gra_0.1} displays the yields of primary, Auger, and
secondary electrons for carbonaceous grains with $a=0.1\micron$.  The 
primary yield $Y_p$ reaches unity at $h\nu = 10^4 \eV$, where both $l_a$ and 
$l_e$ are very large.  However, $Q_{\rm abs} Y_p$ is a decreasing function
at high photon energy.  Figures \ref{fig:y_tot_gra} and 
\ref{fig:y_tot_sil} display the total yield (primary plus Auger plus
secondary) for carbonaceous and silicate grains of various sizes.

\section{Total Photoelectric Emission Rate and Grain Charging}

The primary photoelectric 
emission rate (excluding that due to excess ``attached''
electrons on negatively charged grains) is
\be
\Jpe = \pi a^2 \int_{\nu_{\rm pet}({\rm band})}^{\nu_{\rm max}} d\nu
\frac{cu_{\nu}}{h\nu} \Qabs \left( Y_{p; {\rm band}} + \sum_{i, s}
Y_{p; i,s} \right)
\ee
(c.f. eq.~25 in WD01); $h\nu_{\rm max}$ is the highest photon energy in 
the incident radiation field and $Y_p$ denotes the yield of primary 
electrons.  
For negatively charged grains, the excess electrons undergo photodetachment
at a rate
\be
J_{\rm pd} = \int_{\nu_{\rm pdt}}^{\nu_{\rm max}} d\nu \frac{cu_{\nu}}
{h\nu} \sigma_{\rm pdt}~~~;
\ee
the photodetachment threshold energy and cross section are taken from 
equations 18 and 19 in WD01, respectively.
The Auger electron emission rate is
\be
J_A = \pi a^2 \int_{I_{is, {\rm min}}/h}^{\nu_{\rm max}} d\nu
\frac{cu_{\nu}}{h\nu} \Qabs \sum_{i, s, j}
Y_{A; i,s,j}~~~,
\ee
where $I_{is, {\rm min}}$ is the lowest ionization energy among
those of the inner shells.  Of course, for all shells the yields are zero
when $h\nu < h\nu_{\rm pet}$.  
The emission rate of secondary electrons (excited by primary photoelectrons
and Auger electrons) is
\be
J_{\rm sec, \, pe} = \pi a^2 \int_{\nu_{\rm pet}({\rm band})}^{\nu_{\rm max}}
d\nu \frac{cu_{\nu}}{h\nu} \Qabs \sum_{k} Y_{{\rm sec}, \, k}~~~,
\ee
where the sum over $k$ denotes the sum over all exciting primary and Auger 
electrons.

The collisional charging rates due to positive ions (assumed to have charge
$e$) and electrons are denoted $\Jion$ and $J_e$, respectively, and are 
computed as described in \S \S 2.4 and 3 in WD01.  Since we will be
considering high-temperature gas here, we also include secondary 
emission induced by incident gas-phase electrons.  We compute the charging
rate $J_{\rm sec, \, gas}$ using the Draine \&
Salpeter (1979) model (their eqs.~14 through 16 and A7 through A11).

When each charging event involves the transfer of a single charge 
quantum $\pm e$, statistical equilibrium yields
\be
f_Z(Z) [\Jpe(Z) + J_A(Z) + J_{\rm sec, \, pe}(Z) + J_{\rm pd}(Z) + \Jion(Z)
+ J_{\rm sec, \, gas}] = 
f_Z(Z+1) J_{e}(Z+1)~~~,
\label{eq:balance}
\ee
where $f_Z(Z)$ is the probability for the grain charge to be $Ze$.
WD01 used this equation (without $J_A$, $J_{\rm sec, \, pe}$, and
$J_{\rm sec, \, gas}$) to solve for the grain charge distribution.
When Auger and secondary emission are important, more than one electron can 
be emitted
at a time, yielding a slightly more complicated balance equation and a 
very complicated combinatorics problem.  
To simplify matters, we will continue to use equation (\ref{eq:balance}).
We also compute a single ``equilibrium charge'', i.e., the 
charge for which the net current $\Jpe + J_{\rm pd} + J_A + J_{\rm sec, \, pe}
+ \Jion + J_{\rm sec, \, gas} - J_e$
equals zero.  For large grains, the equilibrium charge is nearly 
identical to the average over the charge distribution.  Even for small
grains, the difference is not large when the grain is highly charged.
In the following sections, we will present results obtained by resolving
the charge distribution, unless otherwise indicated.  

A rigorous treatment of multi-electron ejection events would increase the
value of $Z$ when computing potentials and yields for successive escaping
electrons.  We will ignore this complication as well.  

\section{Gas Heating \label{sec:htg}}

The photoelectric gas heating rate per grain with charge $Ze$, due to 
primary photoelectrons originating in the band structure, is given by
\be
\label{eq:htg_primary_band}
\Gamma_{\rm pe}^{\prime}(a) = \pi a^2
\int_{\nu_\pet({\rm band})}^{\nu_{\rm max}} d\nu \frac{c u_{\nu}}{h\nu} \Qabs 
Y_{p; {\rm band}} 
\int_{E_{\rm min}({\rm band})}^{E_{\rm max}({\rm band})} 
dE f_E({\rm band}; E) E~~~, 
\ee
where $E_{\rm min}$ is given by equation (\ref{eq:E_min}), 
$E_{\rm max} = h\nu - h\nu_\pet + E_{\rm min}$, and 
$f_E(E) = f_E^0(E)/y_2$.  The heating rate due to primary electrons 
originating 
in inner shell ($i$, $s$), $\Gamma^{\prime}_{{\rm pe}; i, s}$, is identical, 
except
that each quantity in equation (\ref{eq:htg_primary_band}) is evaluated for
($i$, $s$).  Similarly for the heating rate due to Auger and secondary 
electrons, except that $E_{\rm max} = I_{i, s} - (Z+1) e^2/a$ for Auger 
emission and secondary emission excited by Auger electrons; $I_{i, s}$ is
the ionization energy of the shell in which the primary excitation occurs.
The heating rate due to photodetachment is given in eq.~40
of WD01.

The gas cooling rate per grain due to recombination of charged particles with
the grain, $\Lambda_{\rm gr}^{\prime}$, is computed as described in \S 5 of
WD01.
The cooling is reduced when secondary electrons 
are ejected.  We make a simple  
estimate of this reduction by assuming that the average energy (with 
respect to infinity) of escaping secondary electrons is 
\be
\bar{E} = \frac{\int_{E_1}^{\langle E_0 \rangle} dE f_E^0(E) (E - e\phi)}
{\int_{E_1}^{\langle E_0 \rangle} dE f_E^0(E)}
\ee
where $\langle E_0 \rangle$ is the average energy of the incident 
electrons (eq.~16 in Draine \& Salpeter 1979), $\phi$ is the grain 
potential, $f_E^0$ is given by
equation (\ref{eq:f_sec}) with $E_{\rm low} = 0$ and $E_{\rm high} = 
\langle E_0 \rangle$, and $E_1 = 0$ when $Z < 0$ and $e\phi$ when 
$Z \ge 0$.

\section{Planetary Nebulae \label{sec:PN}}

A planetary nebula is exposed to a hard radiation field from the hot
central star, and photoelectric emission from dust is an important 
mechanism for heating the gas (Dopita \& Sutherland 2000; van Hoof et
al.~2004).  

In Table \ref{tab:PN}, we display photoelectric heating 
rates $\Gamma_{\rm pe}$ and collisional cooling rates $\Lambda$ for a model 
planetary nebula and
several different grain size distributions.  We assume that the radiation
field is a blackbody with a color temperature $T_c = 2 \times 10^5 \K$
and intensity $G = 2 \times 10^3$. ($G \equiv
\urad^{\rm uv} / \uHab$, where $\urad^{\rm uv}$ is the
energy density in the radiation field between $6 \eV$ and $13.6 \eV$
and $\uHab = 5.33 \times 10^{-14} \erg {\cm}^{-3}$ is the
Habing (1968) estimate of the starlight energy density in this
range.)  This intensity corresponds to a distance of $10^{17} \cm$ from 
a star with luminosity $10^4 \, L_{\odot}$.  
We adopt a fully ionized H nebula with number density 
$\nH = 10^3 \cm^{-3}$ and temperature $T = 10^4 \K$. 

The dust in a planetary nebula is unlikely to be the same as the
     average interstellar dust modeled by Weingartner \& Draine (2001a).
     We assume that either carbonaceous grains or silicate grains are
     present, but not both.  The size distributions for either the
     carbonaceous grains or the silicate grains are taken from the size
     distribution for the appropriate component in various dust models
     developed by Weingartner \& Draine (2001a).  These reproduce extinction
     laws with $R_V=A_V/E(B-V) = 3.1$, 4, or 5.5, for various values of
     the amount $b_C$ of carbon per interstellar
     H atom in PAHs, with cases ``A'' and ``B''
     referring to different assumptions made concerning the total
     amount of material in the interstellar grain model.  The
     photoelectric heating contribution $\Gamma_{\rm pe}/(Gn_{\rm H})$
     of the silicate component varies from 3.6 to
     $23\times10^{-25}\erg\s^{-1}$ -- more than a factor 6 -- showing
     the sensitivity to the actual size distribution.  The heating and
     cooling rates in Table 2 should be taken as an indication of the
     general magnitude that $\Gamma_{\rm pe}$ and $\Lambda$ can have.

For carbonaceous dust, use of the WD01 yields overestimates the heating
rate by a factor of 1.1 to 2 and the cooling rate by as much as 28\%
(for the first entry in Table \ref{tab:PN}).  Inner shell and 
Auger emission are unimportant, and secondary emission contributes only at 
the 5\% level or less.  
For silicate dust, the heating (cooling) rate is overestimated by a factor
of $\approx 1.9$ to 2.5 (1.3 to 1.5) 
when the WD01 yields are employed.  Inner shell
and Auger emission account for about 10\% of the total heating and 
secondary emission contributes at the 5 to 10\% level.  
For both types of dust, the heating (cooling) results differ by at most
5\% (1\%) when secondary emission induced by incident gas-phase electrons
is ignored.  

Since the adopted photoelectric yields are highly uncertain, we also 
computed the heating and cooling rates with the bulk yields increased
and decreased by a factor 2.  For carbonaceous grains, the maximal 
changes in the heating (cooling) rates were $\approx 30\%$ (20\%).  
For silicate grains, the revised yields resulted in changes of factors
$\approx 1.5$ and 0.6 for the heating rate; the cooling rate was somewhat
less sensitive to the changes in the yield.

\section{Quasar Outflows \label{sec:quasar}}

X-ray observations of broad absorption line quasars (BALQSOs) reveal 
large columns of absorbing material (e.g., Gallagher et al.~2002).
Optical observations, on the other hand, indicate that the nucleus is not 
substantially reddened (e.g., Reichard et al.~2003).  Thus, it appears that 
the outflows either are dust-free or contain only very large (by interstellar
standards) grains.  One possible explanation is that the dust is destroyed
by sputtering in the hot gas.  In this process, energetic gas-phase ions 
collide with the grain and eject surface atoms into the gas.  
There are two limiting regimes for sputtering: ``thermal''
     sputtering, where the drift velocity of the grain is negligible
     compared to the thermal motions of the ions, and ``drift''
     sputtering, where the relative velocity of impinging ions is due
     primarily to motion of the grain through the gas.
In the 
presence of a hard radiation field, the grains may acquire large, positive
electric potentials, suppressing destruction by sputtering (Mathews 1967;
Laor \& Draine 1993).  
On the other hand, for sufficiently large potentials, the electrostatic 
stress exceeds the tensile strength of the grain material and the grain 
explodes.    
In this section, we estimate grain potentials and 
lifetimes for conditions likely to characterize BALQSO outflows.

\subsection{Grain Charging}

We adopt the unobscured quasar spectrum from Sazonov et al.~(2004), given
by
\be
\label{eq:sazonov}
u_E = \alpha \cases{
0.0798 (E/\eV)^{-0.6} &, $1 \eV \le E < 10 \eV$\cr
(E/\eV)^{-1.7} \exp(E/2 \, {\rm keV}) &, $10 \eV \le E < 2 \, {\rm keV}$\cr
2.94 \times 10^{-3} (E/\eV)^{-0.8} \exp(-E/200 \, {\rm keV}) &, 
$E \ge 2 \, {\rm keV}$\cr
}~~~.
\ee
The normalization constant $\alpha = 3.07 \eV^{-1} u(> 13.6 \eV) = 
2.26 \times 10^{-10} \erg \eV^{-1} n_{\gamma}(> 13.6 \eV)$, 
where $u(> 13.6 \eV)$ [$n_{\gamma}(> 13.6 \eV)$]
is the total energy density [photon number density] beyond $13.6 \eV$.
We consider gas temperatures $T = 10^4$, $10^5$, and $10^6 \K$ and 
ionization parameters $U \equiv n_{\gamma}(> 13.6 \eV)/\nH = 0.1$, 1, 10, 
and $10^2$.  

\begin{figure}
\plottwo{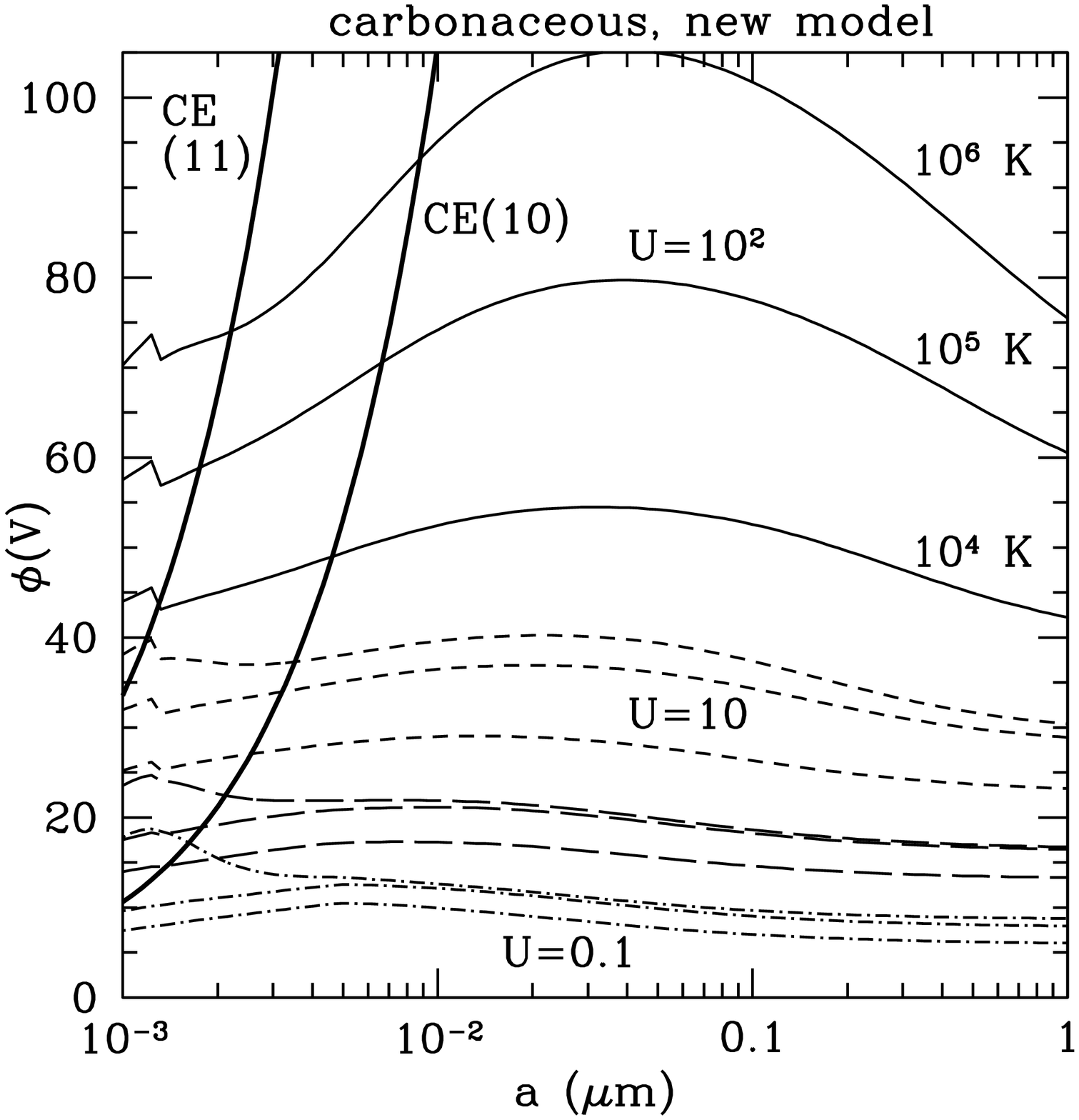}{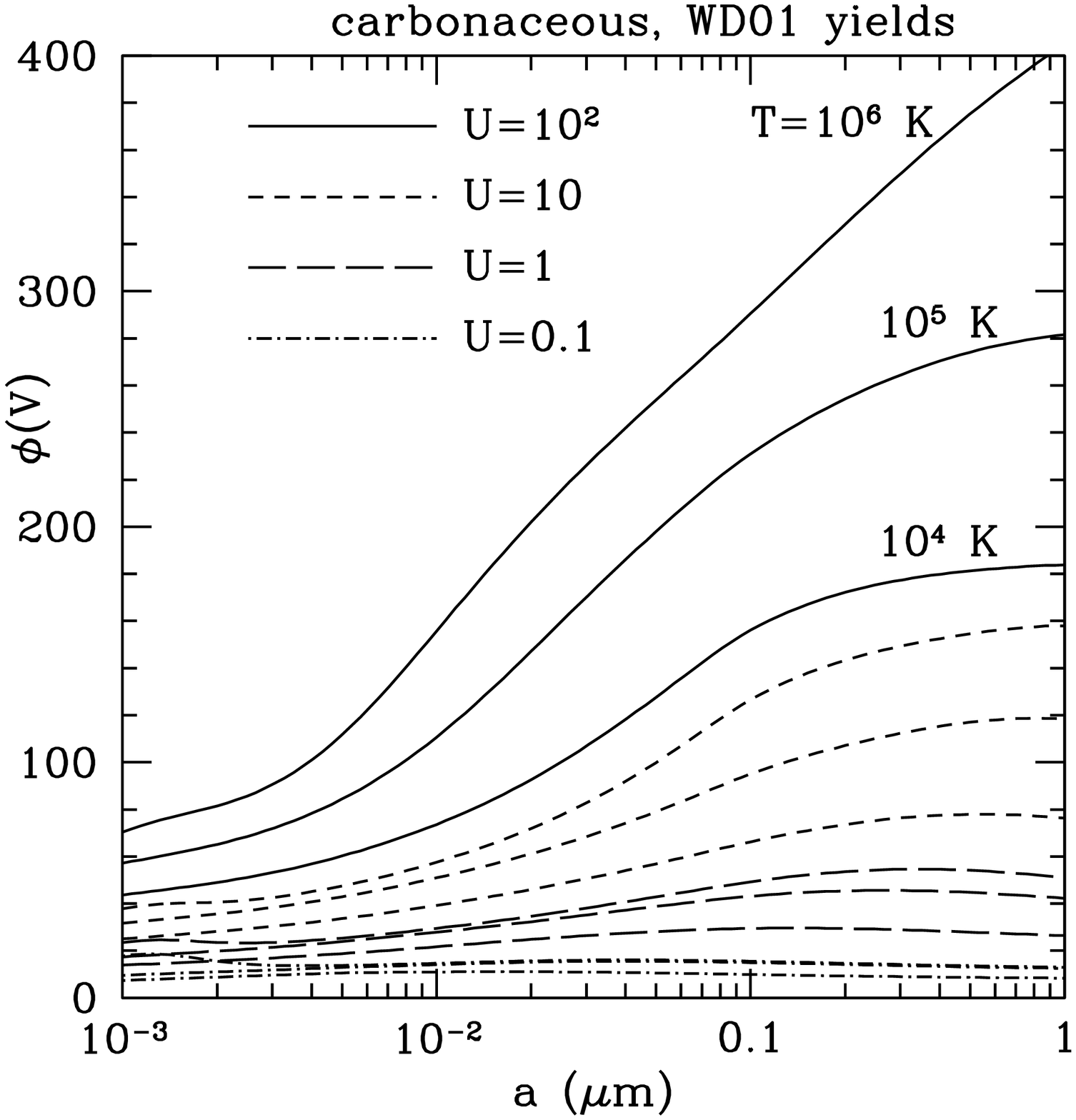}
\caption{
\label{fig:phi_quasar_gra}
(a) Carbonaceous grain potential $\phi$ versus radius $a$ for grains exposed
to an unobscured quasar spectrum (eq.~\ref{eq:sazonov}) with ionization 
parameter $U=0.1$, 1, 10, and $10^2$, and 
for gas temperatures $T=10^4$, $10^5$, and $10^6 \,$K, as indicated.
For a given value of $U$, the potential is larger for higher $T$.
Coulomb explosions occur to the left of the 
heavy curve labelled ``CE(10)'' if the tensile strength 
$S_{\rm max} \approx 10^{10} \, {\rm dyn} \cm^{-2}$ and to the left of 
the curve labelled ``CE(11)'' if 
$S_{\rm max} \approx 10^{11} \, {\rm dyn} \cm^{-2}$.
(b) Computed grain potentials using the photoelectric yields from WD01, 
for comparison. 
        }
\end{figure}

\begin{figure}
\plottwo{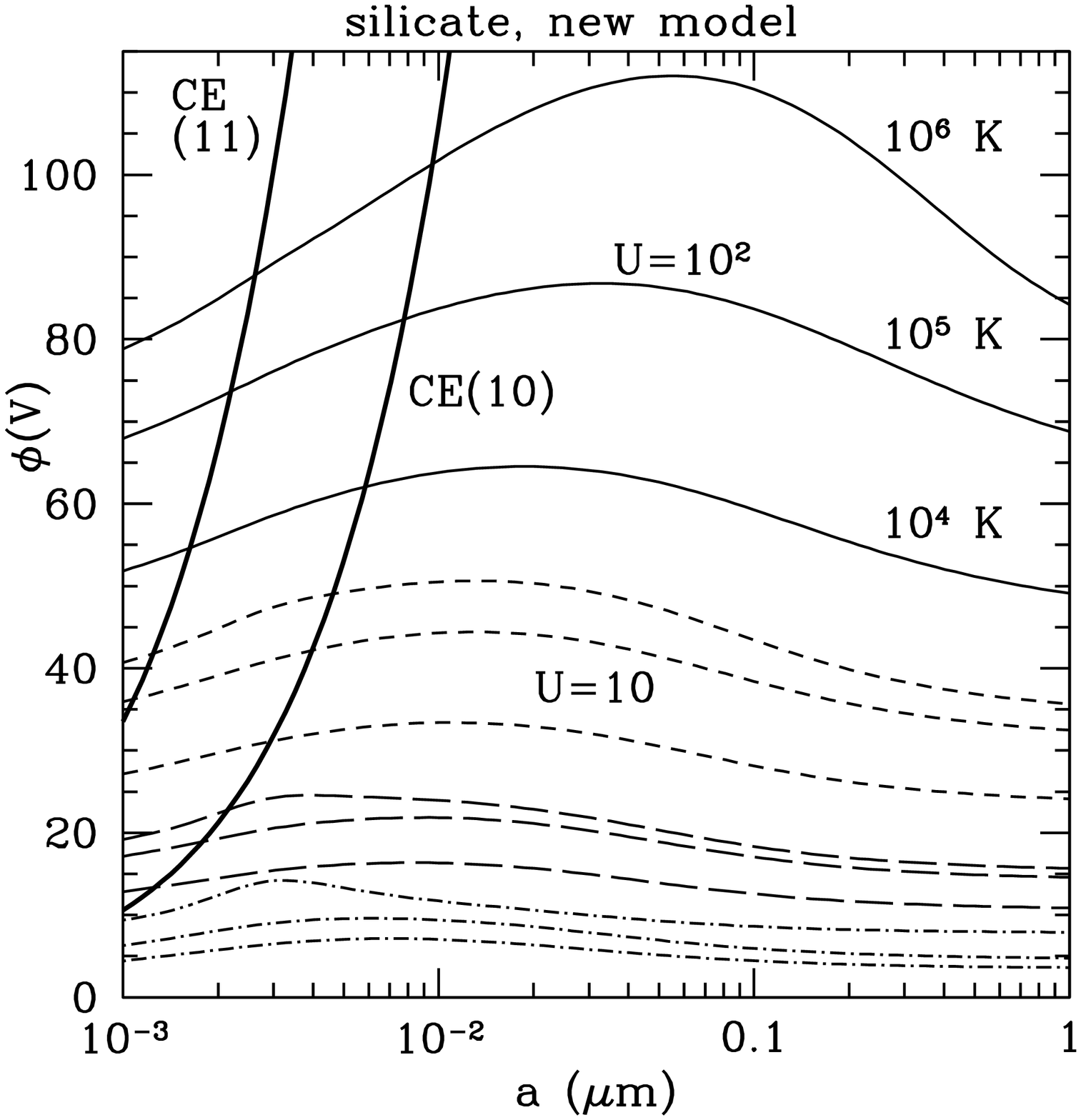}{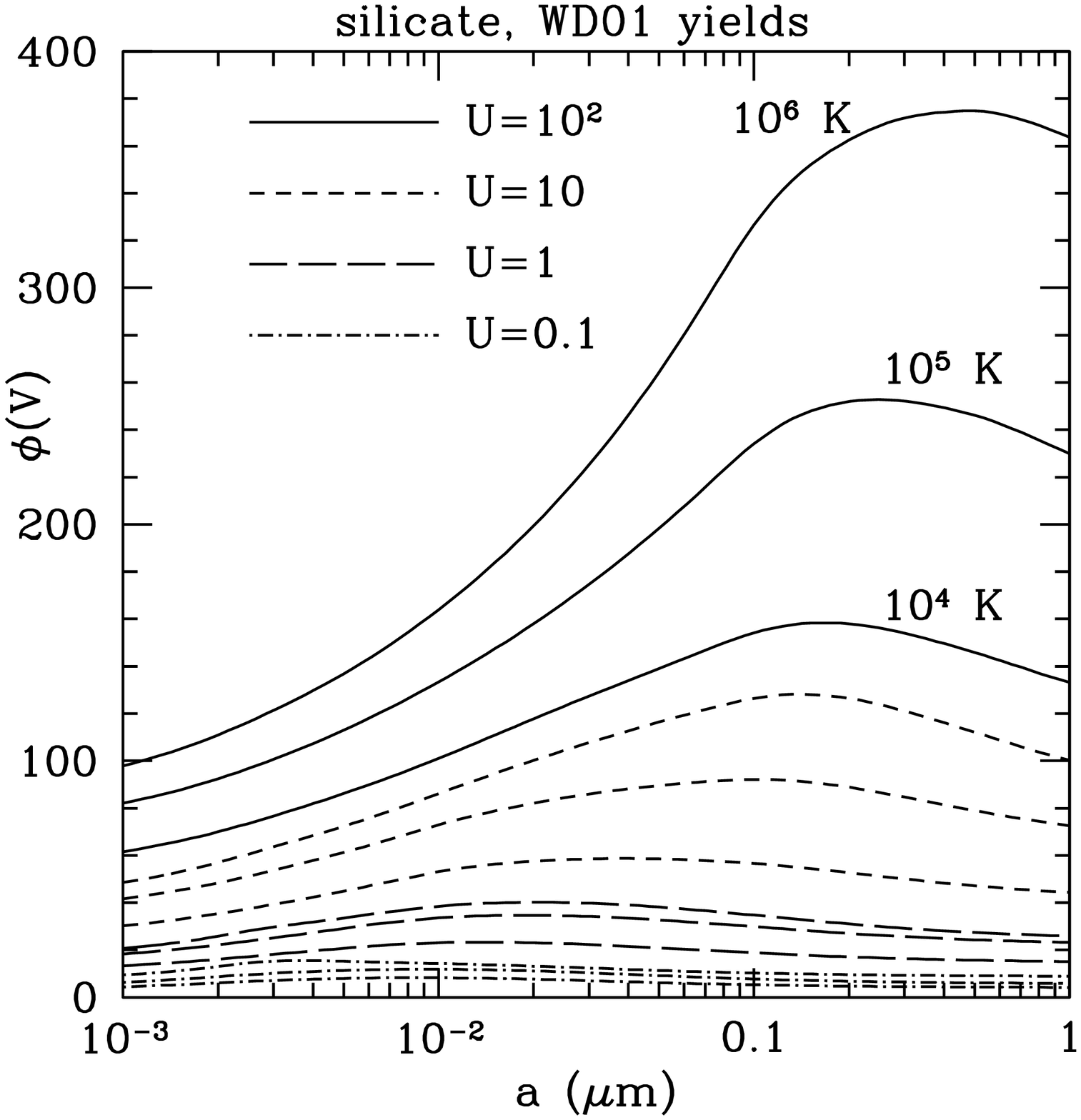}
\caption{
\label{fig:phi_quasar_sil}
Same as Fig.~\ref{fig:phi_quasar_gra}, except for silicate grains.
        }
\end{figure}

Figures \ref{fig:phi_quasar_gra}a and \ref{fig:phi_quasar_sil}a 
display the grain potential $\phi$ for carbonaceous and silicate grains as 
a function of size.  For comparison, Figures 
\ref{fig:phi_quasar_gra}b and \ref{fig:phi_quasar_sil}b display the
potentials computed using the WD01 photoelectric yields; in this case,  
we only include photoelectric emission from 
the band structure, photodetachment, and collisional charging (including 
secondary emission induced by colliding gas-phase electrons).  
The use of our new, more realistic yields 
for photoelectric emission from the band structure substantially reduces the 
computed potential, despite the additional charging due to the 
emission of inner shell and Auger electrons (which account for at most
$\approx 30$\% of the charging) and the emission of photo-induced 
secondaries (which account for at most 2\% of the charging).  
Secondary emission induced by gas-phase electrons contributes at most at
the 5, 10, and 25\% levels for $U=10^2$, 10, and 1, respectively.  For
$U=0.1$ and $T=10^6 \K$, this process dominates the grain charging.
The ``notches'' at the far left in Figure \ref{fig:phi_quasar_gra}a arise
because we adopt higher yields for photoelectric emission from the band
structure of carbonaceous grains with $a < 13 \Angstrom$ 
(eq.~\ref{eq:y_band_PAH}).
To test how sensitive the results are to uncertainties in the 
photoelectric yield, we computed the potentials for the case that 
$U = 10$ and $T = 10^5 \K$ but with the adopted bulk yields multiplied by a 
factor of 2 or 0.5.  The largest changes in the potential were by factors
1.3 and 0.75.

\subsection{Grain Destruction}

\subsubsection{Coulomb Explosions}

The electrostatic stress on a spherical grain is $S = (\phi/a)^2/4\pi$;
if $S$ exceeds the tensile strength $S_{\rm max}$, then the grain
fragments (Draine \& Salpeter 1979; Waxman \& Draine 2000; Fruchter 
et al.~2001).   
Real grains are not perfectly spherical and the structure
of grain material is poorly known; thus $S_{\rm max}$ is highly
uncertain.  The maximum grain potential
for which fragmentation will not occur is
\be
\phi_{\rm max} = 1.06 \times 10^3 \, {\rm V} \left( \frac{S_{\rm max}}
{10^{10} \, {\rm dyn} \cm^{-2}} \right)^{1/2} \left( \frac{a}{0.1 \micron}
\right)~~~.
\ee
If $S_{\rm max} \approx 10^{11} \, {\rm dyn} \cm^{-2}$, as measured for
ideal materials, then ion field emission, rather than Coulomb explosions,
may limit the positive grain charge (Waxman \& Draine 2000).  

Figures \ref{fig:phi_quasar_gra}a and \ref{fig:phi_quasar_sil}a
show the locus for Coulomb explosions for two possible
     values $S_{\rm max}$ of the tensile strength.
     The results in Figures 
\ref{fig:phi_quasar_gra}a and \ref{fig:phi_quasar_sil}a
 show that if
     $S_{\rm max}\approx 10^{10} \, {\rm dyn} \cm^{-2}$, then for ionization
     parameter $U \gtsim 10$, Coulomb explosions result in a minimum
     grain radius of several $\times 10^{-3}\micron$.  However, electrostatic
     stresses reach $10^{11} \, {\rm dyn} \cm^{-2}$ 
only for the smallest grains and very high values of $U$.

\subsubsection{Thermal Sputtering}

The cross section for a collision between an ion with charge $z_i e$
and energy $E_i$ and a grain with radius $a$ and potential $\phi$ is
\be
\sigma = \pi a^2 \times \cases{1 - z_i e \phi /E_i &, $E_i > z_i e \phi$\cr
0 &, $E_i \le z_i e \phi$\cr}~~~.
\ee
Note that we have neglected the polarization of the grain by the ion.
Integrating over a Maxwell speed distribution and summing over the various
ions in the gas yields the sputtering rate for the case that $\phi \ge 0$:
\be
\frac{1}{\nH} \frac{da}{dt} = \frac{2 \pi \mu m_p}{\rho} (2 \pi k T)^{-3/2}
\sum_i \frac{n_i}{\nH} m_i^{-1/2} \int_{z_i e \phi}^{\infty} dE_i \, E_i
\, \exp(-E_i/kT) \left( 1 - \frac{z_i e \phi}{E_i} \right) Y_{s,i}(E_i -
z_i e \phi)~~~,
\ee
where $\rho$ is the mass density of the grain material, 
$\mu m_p$ is the average mass of the grain consituent atoms,
$m_i$ is the mass of ion $i$, $n_i$ is the
number density of ion $i$, and $Y_{s,i}$ is the sputtering yield for ion
$i$ (i.e., the probability that an atom from the grain is ejected following
a collision with a gas-phase ion).
For the case that all of the ions are singly ionized,
\be
\label{eq:sputtering_rate}
\frac{1}{\nH} \frac{da}{dt} = \exp(-e \phi/kT)
\frac{1}{\nH} \left. \frac{da}{dt} \right|_{\phi =0}~~~.
\ee

For the Sazonov et al.~(2004) spectrum (eq.~\ref{eq:sazonov}), 
\be
\label{eq:n_gamma_sazonov}
n_{\gamma}(> 13.6 \eV) = 2.00 \times 10^7 \cm^{-3} 
\left( \frac{L}{10^{46} \erg 
\s^{-1}} \right) \left( \frac{r}{\rm pc} \right)^{-2}~~~,
\ee
where $L$ is the luminosity of the quasar between 1$\,$eV and 10$\,$keV
and $r$ is the distance to the central source.
Combining equations (\ref{eq:sputtering_rate}) and 
(\ref{eq:n_gamma_sazonov}) yields the following expression for the  
lifetime of a grain with radius $a$ against thermal sputtering:
\be
\tau_{\rm ts} = 5.00 \yr \left( \frac{a}{0.1 \micron} \right) \left(
\frac{U}{10} \right) \left( \frac{L}{10^{46} \erg \s^{-1}} \right)^{-1}
\left( \frac{r}{\pc} \right)^2 \left( \frac{n_{\rm H}^{-1} da/dt (\phi =0)}
{-10^{-4} \cm^3 \Angstrom \yr^{-1}} \right)^{-1} \langle \exp(e \phi /kT)
\rangle_a~~~,
\ee
where
\be
\langle \exp(e \phi /kT) \rangle_a \equiv \frac{1}{a}
\int_{a_{\rm min}}^a da^{\prime}
\exp[e\phi(a^{\prime})/kT]~~~.
\ee
We take $a_{\rm min} = 5 \Angstrom$ and use the equilibrium charge for 
each grain size, rather than summing over the charge distribution.  
Tielens et al (1994) reexamined thermal sputtering as a function
     of gas temperature, obtaining rates very similar to those found by
     Draine \& Salpeter (1979).
We adopt the approximate expression from Tielens et al.~(1994) for the 
thermal sputtering rate for neutral grains (their eq.~4.21 and Table 4);
these were derived 
assuming that H, He, C, N, and O are all singly ionized and have solar
abundances.  

\begin{figure}
\epsscale{1.00}
\plotone{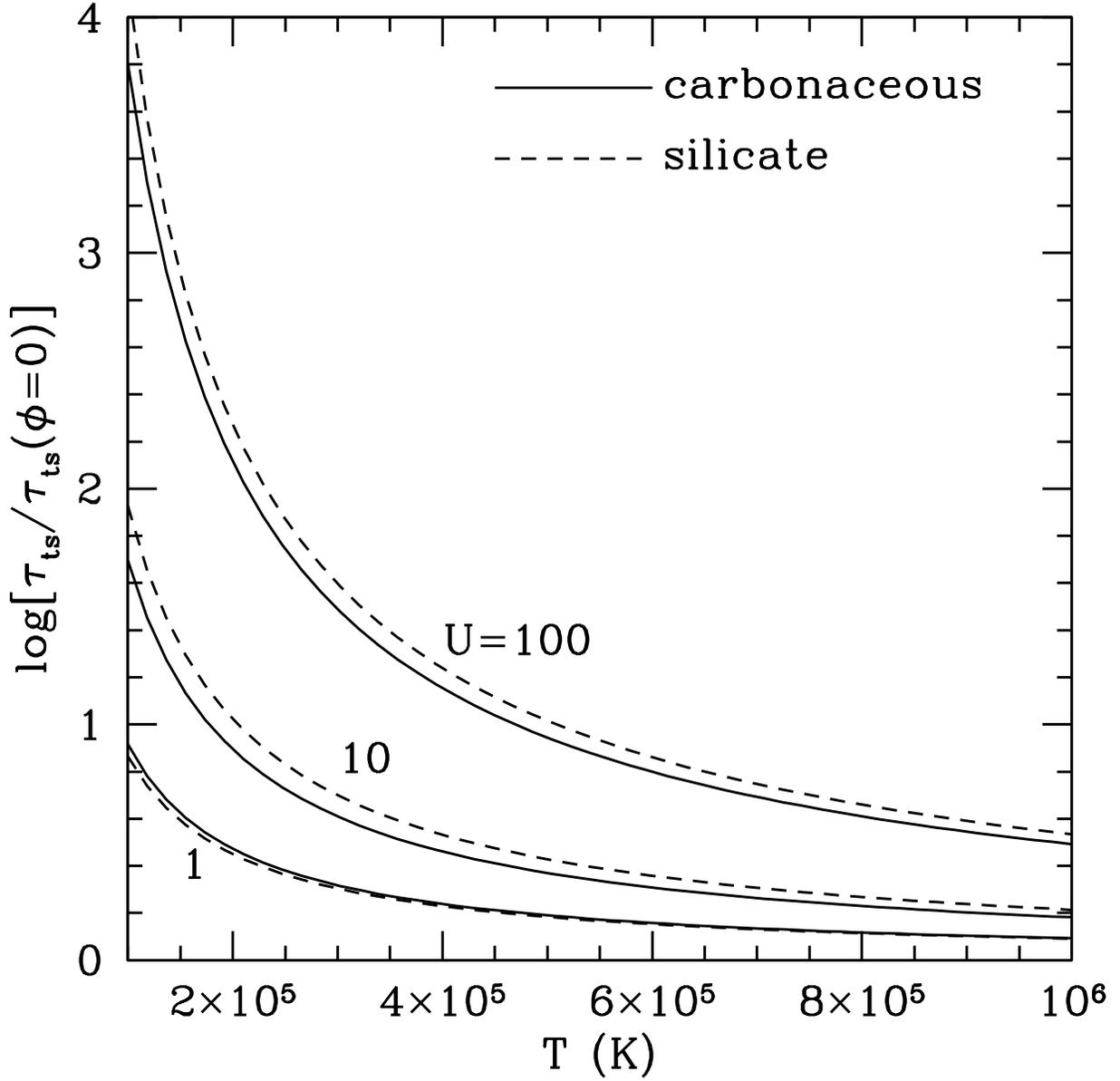}
\caption{
\label{fig:quasar_enhance}
Ratio of the thermal sputtering timescale to its value for uncharged grains,
versus gas temperature.
        }
\end{figure}

\begin{figure}
\epsscale{1.00}
\plotone{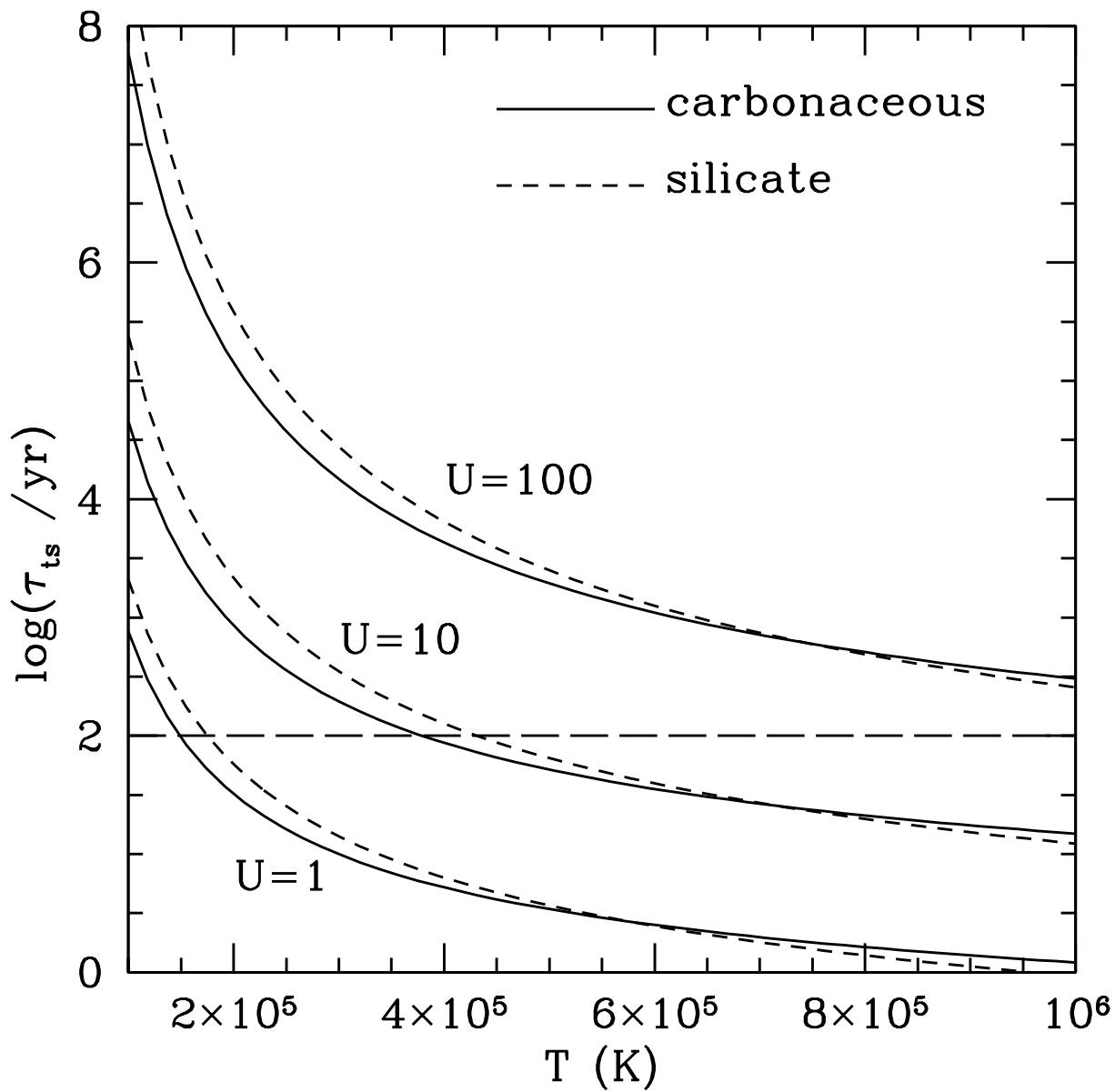}
\caption{
\label{fig:quasar_lifetimes}
Thermal sputtering lifetime $\tau_{\rm ts}$
for grains with $a=0.3 \micron$ when $L = 10^{46} \erg \s^{-1}$
and $r = 3 \pc$, versus gas temperature.  
The horizontal long-dashed line is the
outflow timescale $\tau_{\rm flow}$, assuming that the outflow speed
$v \sim 3 \times 10^4 \kms$.
        }
\end{figure}

Figure \ref{fig:quasar_enhance} displays the ratio 
$\langle \exp(e \phi /kT) \rangle_a$ of $\tau_{\rm ts}$ to
the value it would take if the grains were uncharged.  
Figure \ref{fig:quasar_lifetimes} presents thermal sputtering lifetimes
for grains with $a=0.3 \micron$ when $L = 10^{46} \erg \s^{-1}$ 
and $r = 3 \pc$.  The horizontal long-dashed line is the
outflow timescale $\tau_{\rm flow} = r/v \sim 
100 \yr (r/3 \pc)$ if the flow speed $v \sim 3 \times 10^4 \kms$.
The grains survive only for quite large values of the ionization parameter
or low values of the gas temperature.  Furthermore, when $U \gtsim 10$,
grains with $a \approx 0.3 \micron$ experience supersonic drift relative
to the gas, increasing the destruction rate.  
Thus, the lack of nuclear reddening in BALQSOs does not appear to require
that the outflow originates in a dust-free region (e.g., within the 
sublimation radius).  However, better observational constraints on the 
location and physical conditions in the outflow are needed before a firm 
conclusion can be reached.

\section{Summary}

In this paper, we have extended the WD01 photoelectric emission model to higher
photon energies, enabling the treatment of dust exposed to hard radiation 
fields.  We provide revised yields for photoelectric
emission from the band structure
when $h\nu > 20 \eV$ and treat the emission of Auger electrons and 
secondary electrons excited by primary and Auger electrons.  

For H$\,$II regions, ionized by OB stars, the radiation field is soft
enough that the WD01 model can be employed without modification; the
new treatment yields  a $\ltsim 10\%$ difference in the gas heating rate.

For planetary nebulae, the gas heating rates found with the extended model
are lower, by a factor 2 or less, 
than those estimated with the WD01 model, since 
the revised yields for photoelectric 
emission from the band structure are lower.  
Inner shell, Auger, and secondary emission are not important.  
For quasar outflows, 
the new treatment yields significantly lower grain potentials.  In this
case, combined inner shell and Auger emission contribute at as high as 
the 30\% level.  This is less than the uncertainty associated with 
photoelectric emission from only the band structure.  Secondary emission 
is unimportant.  It appears that thermal sputtering can destroy dust in 
BALQSO outflows in less than a dynamical time, for a wide range of 
plausible physical conditions in the outflow.  However, better constraints
on the outflow conditions are needed to definitively settle the issue of
dust survivability.

In environments with harder radiation fields (e.g. supernova remnants
and the central regions of galaxy clusters), inner shell, Auger, and 
secondary emission may be more important relative to photoelectric emission 
from the band structure.  However, in these cases the gas is so hot
that secondary emission induced by gas-phase electrons and ions 
incident on the grain is expected to dominate over photoelectric emission in 
grain charging (Draine \& Salpeter 1979).

\acknowledgements

Support for this work, part of the Spitzer Space Telescope Theoretical 
Research Program, was provided by NASA through a contract issued by the 
Jet Propulsion Laboratory, California Institute of Technology under a 
contract with NASA.  BTD was supported in part by NSF grant AST-0406883.
We are grateful to Norm Murray, Gary Ferland, and Peter van Hoof for helpful 
discussions.

\begin{deluxetable}{llllllll}
\tablewidth{0pc}
\tablecaption{Ionization Energies\tablenotemark{a}
\label{tab:IE}}
\tablehead{
\colhead{Element}&
\colhead{1s}&
\colhead{2s}&
\colhead{2p}&
\colhead{3s}&
\colhead{3p}&
\colhead{3d}&
\colhead{4s}
}
\startdata
C  &291  &19.39  &11.26 &...   &...   &...   &...\\
O  &538  &28.48  &13.62 &...   &...   &...   &...\\
Mg &1311 &94.0   &54.9  &7.646 &...   &...   &...\\
Si &1846 &156    &106   &15.17 &8.152 &...   &...\\
Fe &7124 &857    &724   &104   &66    &14.7  &7.902\\
\enddata
\tablenotetext{a}{in eV}
\end{deluxetable}

\begin{deluxetable}{llllll}
\tablewidth{0pc}
\tablecaption{Photoelectric Heating and Recombination Cooling Rates for a
Model Planetary Nebula \label{tab:PN}}
\tablehead{
\colhead{composition}&
\colhead{$R_V$}&
\colhead{$10^5 b_{\rm C}$}&
\colhead{case}&
\colhead{$\Gamma_{\rm pe}/(G \nH)$\tablenotemark{a}}&
\colhead{$\Lambda/(G\nH)$\tablenotemark{a}}
}
\startdata
carbon    &3.1  &0.  &A   &16.2  &1.97\\
carbon    &3.1  &6.  &A   &45.9  &5.56\\
carbon    &4.0  &0.  &A   &12.8  &1.55\\
carbon    &4.0  &4.  &A   &31.6  &3.82\\
carbon    &5.5  &0.  &A   &9.75  &1.19\\
carbon    &5.5  &3.  &A   &23.4  &2.84\\
carbon    &4.0  &0.  &B   &16.7  &2.05\\
carbon    &4.0  &4.  &B   &32.0  &3.89\\
carbon    &5.5  &0.  &B   &13.3  &1.64\\
carbon    &5.5  &3.  &B   &22.6  &2.75\\
silicate  &3.1  &0.  &A   &15.6  &1.29\\
silicate  &3.1  &6.  &A   &22.5  &1.89\\
silicate  &4.0  &0.  &A   &9.84  &0.818\\
silicate  &4.0  &4.  &A   &9.31  &0.775\\
silicate  &5.5  &0.  &A   &3.83  &0.314\\
silicate  &5.5  &3.  &A   &3.61  &0.297\\
silicate  &4.0  &0.  &B   &10.1  &0.840\\
silicate  &4.0  &4.  &B   &9.73  &0.810\\
silicate  &5.5  &0.  &B   &3.87  &0.316\\
silicate  &5.5  &3.  &B   &3.84  &0.314\\
\enddata
\tablenotetext{a}{$10^{-25} \erg \s^{-1}$}
\end{deluxetable}

\end{document}